\documentclass[twocolumn,amsmath,amssymb,superscriptaddress,aps,pra]{revtex4-1}

\usepackage{graphicx}
\usepackage{dcolumn}
\usepackage{bm}

\begin{document}

\title{An all-optical nanomechanical heat engine}

\author{Andreas Dechant}
\affiliation{Department of Physics, Friedrich-Alexander-Universit\"at Erlangen-N\"urnberg, 91058 Erlangen, Germany}
\author{Nikolai Kiesel}
\affiliation{Vienna Center for Quantum Science and Technology (VCQ), Faculty of Physics, University of Vienna, A-1090 Vienna, Austria}
\author{Eric Lutz}
\affiliation{Department of Physics, Friedrich-Alexander-Universit\"at Erlangen-N\"urnberg, 91058 Erlangen, Germany}

\begin{abstract}
We propose and theoretically investigate a nanomechanical heat engine. We show how a levitated nanoparticle in a harmonic optical trap inside an optical cavity can be used to realize  a Stirling cycle in the underdamped regime. The all-optical approach enables fast and flexible control of all the thermodynamical parameters and the efficient optimization of the performance of the engine. We  develop a systematic optimization procedure to determine optimal driving protocols. We further perform numerical simulations with realistic parameters and evaluate the maximum power and the corresponding efficiency.

\end{abstract}


\maketitle

During the last decade significant progress in the fabrication of mechanical devices at the micro- and nanoscale has been achieved \cite{cer09}. While this development enabled a plethora of technological applications, it also allows new experiments at the foundations of modern physics. On the one hand, these devices operate in a regime where thermal fluctuations are relevant, which requires appropriate theoretical tools. In this direction,  research in stochastic thermodynamics has been very successful at extending the laws of macroscopic thermodynamics to the level of single trajectories  \cite{sek10,Seifert2012}. The discovery  of fluctuation theorems has further opened the way to a systematic investigation of  far-from-equilibrium processes \cite{jar11,cil13}. On the other hand, experiments with nano- and micromechanical objects have entered the quantum domain, where quantum fluctuations are dominant. Important examples are the recent achievements of optomechanical cavity cooling of oscillators to the ground state \cite{OConnell2010, Teufel2011, Chan2011} and the  experimental demonstration of quantum state preparation \cite{OConnell2010} and entanglement generation \cite{Palomaki2013}.

A paradigmatic system for the study of stochastic thermodynamics are optically trapped micro- and nanobeads \cite{Seifert2012,jar11,cil13}. The optical tweezer allows a fast control of the potential landscape experienced by the particles and an accurate  recording of their trajectories. In a pioneering experiment, this approach has  been used to demonstrate a classical micromechanical Stirling engine \cite{Blickle2011}, where the temperature of the liquid heat bath of the microparticle was controlled by laser absorption. While this is a very natural environment, there are limitations imposed on the accessible parameter regime for the temperature of the liquid and for the optimization of the protocols employed to implement the thermodynamic cycle. 

The future realization of quantum heat engines requires the investigation of  much more isolated systems. Towards this end, a concrete experiment  to build an Otto heat engine using  a single ion in a Paul trap has  been put forward  \cite{Abah2012,ros14}. This ion is completely isolated from its natural   environment, which is substituted by a reservoir of light that is engineered via Doppler cooling. More recently, a scheme to realize an  optomechanical quantum heat engine that operates on polariton modes in the strong coupling regime has been suggested \cite{Zhang2014}. 

In this paper, we propose a levitation approach to nanomechanical heat engines. Submicron particles are here optically trapped in a moderate vacuum in a harmonic potential with variable frequency.  The heat bath is provided by a thermal environment (constituted by the rest gas inside the trap) in combination with optomechanical cavity cooling \cite{Marquardt2007, Genes2008a, Wilson-Rae2008}, that leads to additional tunable damping. To operate the engine, the motion of the particle is underdamped and only weakly coupled to the optical cavity. Altogether, this approach combines the excellent control offered by optical trapping with the fast optomechanical control of the center-of-mass temperature provided by cavity cooling. This allows flexibility in optimizing the heat engine  and gives access to a large parameter regime for the temperature, in principle down to the quantum ground state. Optimization is an essential tool to maximize the performance of a machine given existing constaints \cite{kir98}. In the overdamped regime, the thermodynamic optimization problem  has been solved for harmonic \cite{sch07} and nonharmonic \cite{aur11} systems. Explicit optimal protocols have been obtained for the  Carnot cycle \cite{sch08}, but, to our knowledge, never implemented experimentally. By contrast, optimization in the underdamped case is notoriously more difficult \cite{gom08}, owing to the larger parameter space, and has been little explored. In the following, we begin by describing the working principles of an all-optical optomechanical heat engine. We develop  theoretical methods to analyze and optimize the stochastic engine in the underdamped regime. We present a systematic procedure to determine the driving protocol that maximizes the power output  and evaluate the corresponding efficiency at maximum power. We finally numerically simulate the operation of the engine for a state-of-the art levitated optomechanical system \cite{Kiesel2013} and discuss the occurrence of jumps in the optimal  protocols.

\begin{figure*}[ht]
 \includegraphics[width=0.98\textwidth, clip, trim=0cm 0cm 0cm 0cm]{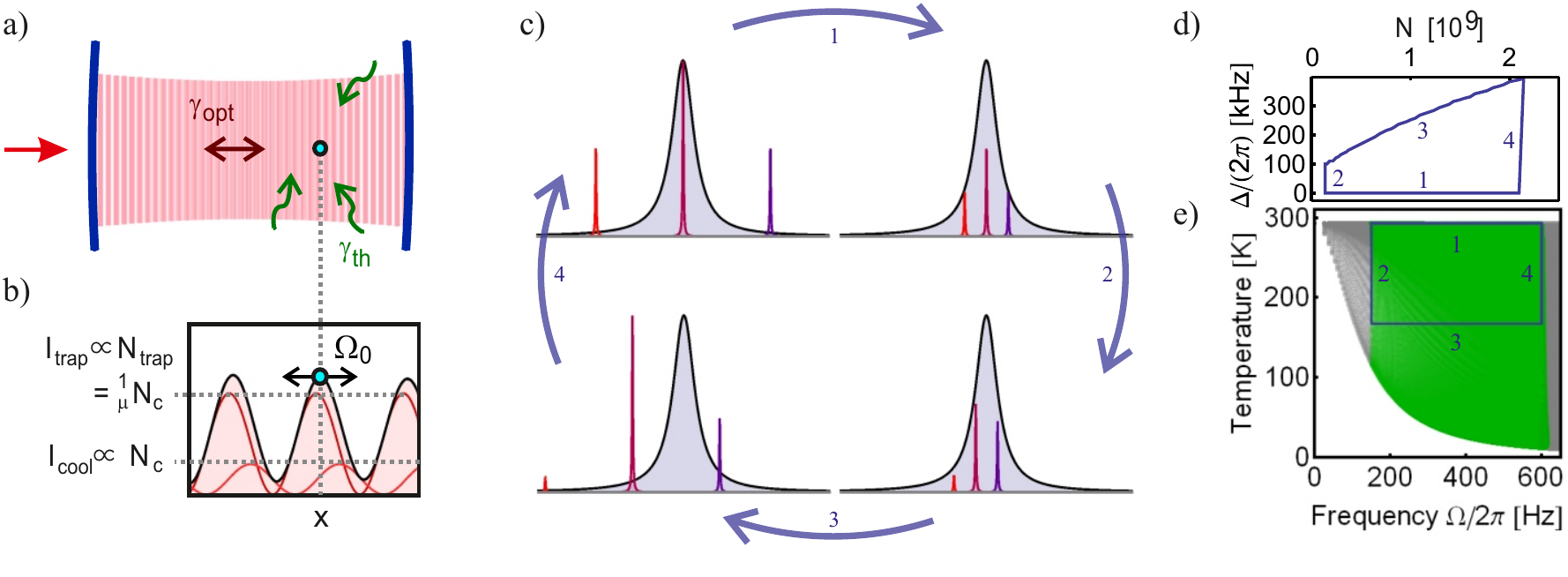}
  \caption{Implementation of a Stirling cycle by cavity cooling of a levitated particle. a) The nanoparticle is optically levitated in the optical field inside a Fabry-Perot cavity that is driven by two light fields (control and trapping beam). By collisions with air molecules its center-of-mass motion is coupled to a thermal environment at room temperature with a damping rate $\gamma_{\text{th}}$. Additional damping $\gamma_{\text{opt}}$ is provided by optomechanical coupling. b) The combined intensity distribution of the two beams defines the optical trap position and frequency $\Omega_0$. Varying the population of the two cavity modes yet keeping their ratio constant ($\mu=N_\text{c}/N_\text{trap}$) allows to vary the mechanical frequency without moving the trap position. Optomechanical coupling to the control mode is ensured by a phase shift between the two modes at the particle position. c) Cooling is achieved by an enhanced scattering of photons into the blue sideband when the cooling beam is red detuned from cavity resonance. In step 3 (compression in cold bath) the detuning is  properly adjusted to keep a constant cooling rate (and thus temperature) while frequency of the mechanical motion is increased. The parameter set required for the whole cycle is shown in d). e) The resulting Stirling heat engine cycle consists of two isothermal and two isochoric processes. As an example, we restrict the accessible frequencies (green area) to values that can directly be achievedare easily accessible in a recent experiment \cite{Kiesel2013}. Note that a wider range should be easily accessible. The colored region represents the accessible parameters assuming the values for thermal and optomechanical coupling achieved in that experiment (see table Tab.~\ref{tab1}). Within these boundaries the Stirling cycle is optimized.}
   \label{fig:cycle}
\end{figure*}

\textit{Optomechanical heat engine.} We consider a nanoparticle trapped in the optical cavity shown in Fig.~1a. In our scenario the particle is well-confined inside the optical trap such that its motion $x(t)$ can be described by a harmonic oscillator of frequency $\Omega_0$ in a thermal environment. 
We describe the system by the following effective Langevin dynamics:
\begin{equation}
\ddot{x}+(\gamma_{\text{th}}+\gamma_{\text{opt}}) \dot{x} + \Omega^2 x = F_{\text{th}} / m.
\label{eq:oscillator}
\end{equation}
Here $\gamma_{\text{th}}$ ($\gamma_{\text{opt}}$) denotes the damping coefficient due to the surrounding gas (sideband cooling), $\Omega$  the effective frequency of the harmonic oscillator, $m$ the mass of the nanoparticle and $F_{\text{th}}$ a delta-correlated noise force  generated by the  collisions with the surrounding gas at temperature $T$. 
For typical experimental parameters (see Tab.~\ref{tab1}), we are in the underdamped regime, $\gamma_\text{eff} = \gamma_{\text{th}}+\gamma_{\text{opt}} < \Omega$, where the oscillatory dynamics of the levitated particle is faster than the thermal equilibration. The steady temperature of the particle is given by $T_\text{eff} = T \gamma_\text{th}/\gamma_\text{eff} < T$.

The proposed experimental scheme for the realization of the  engine is sketched in Fig.~\ref{fig:cycle} a and b. A submicron particle is optically trapped at the intensity maximum of two standing waves in the cavity field whose intensity maxima are shifted in position with respect to each other. One of the fields (control field) serves for cavity cooling and can be detuned from the cavity resonance for that purpose. The other field (trapping field) is kept resonant with the cavity and does not directly participate in the optomechanical interaction, but serves to control the spring constant $\Omega^2$. In this self-trapping approach, that has been experimentally demonstrated in Ref.~\cite{Kiesel2013}, the optomechanical damping $\gamma_{\text{opt}}$ is given by the power $P_\text{c}$ and the detuning $\Delta_\text{c}$ of the control field and the frequency bare $\Omega_0$ of the mechanical resonator. We assume that any noise introduced by these light fields is negligible compared to the thermal noise. The optomechanical coupling between control field and nanoparticle can not only be used to manipulate, but also to detect its motion. Specifically, the axial motion of the particle generates a phase modulation of the control field which can be detected by heterodyne detection \cite{Kiesel2013}.

The frequency $\Omega_0$ of the mechanical resonator is determined by the intracavity power of the cooling ($P_\text{c}$) and the trapping beam ($P_\text{t}$). In addition, the frequency is also modified by the optical spring effect, which results in a shift of the mechanical frequency $\Omega_0\rightarrow \Omega$ and depends on the detuning $\Delta_\text{c}$ \cite{Genes2008a}. Note that the optomechanical cooling may result in a deviation from the thermal equilibrium state for the mechanical oscillator, this effect, however, is negligible for the parameter regime we discuss here. For simplicity, we want to ensure that the position of the optical trap stays fixed, which can be achieved by keeping the ratio of the photon occupation of the cooling and trapping mode ($N_{\text{c}}$ and $N_{\text{trap}}$) fixed to $\mu=\frac{N_{\text{c}}}{N_{\text{trap}}}$, choosing the powers $P_\text{t}$ and $P_\text{c}$ accordingly.

The temperature of the bath may be regulated via sideband cooling. 
For the presented experimental configuration the theory of sideband cooling  for  standard clamped optomechanics \cite{Marquardt2007, Genes2008a, Wilson-Rae2008, Chang2010, Romero-Isart2010, Barker2010, Kiesel2013} directly applies. In this scheme, the oscillating particle scatters photons into optical sidebands of frequencies $\omega_{\text{c}}\pm\Omega_{0}$ at rates $A_{\pm}=\frac{1}{4}\frac{g_{0}^{2}\langle\hat{n}\rangle\kappa}{(\kappa/2)^{2}+(\Delta_\text{c}\pm\text{\ensuremath{\Omega_{0}}})^{2}}$, known as Stokes and anti-Stokes scattering, respectively. The parameter $\kappa$ denotes the FWHM cavity linewidth, $\langle\hat{n}\rangle$ the thermal photon number and $g_0$ the optomechanical single photon coupling. For $\Delta_\text{c}>0$ (red detuning), anti-Stokes scattering becomes resonantly enhanced by the cavity. This process results in a damping of the center-of-mass motion of the particle with an additional friction coefficient  $\gamma_\text{opt}=A_{-}-A_{+}$ that can be easily varied via $\Delta_\text{c}$. Note that variations of the frequency $\Omega$ due to the optical spring effect can be compensated by adapting the intracavity field. 

The two control parameters of the optomechanical heat engine, the frequency $\Omega$ and the optical damping $\gamma_{\text{opt}}$ (which sets the effective temperature $T_\text{eff}$), can thus be directly tuned via the two control parameters of the experiment,   the depth of the optical trap (via $N_\text{c}$) and the detuning $\Delta_\text{c}$. A Stirling cycle that consists of two isochoric and two isothermal transformations may then be implemented in the following way:

\noindent \textit{Step 1:} The particle interacts with a bath at constant temperature $T$ via the coupling $\gamma_{\text{th}}$ (both laser fields are resonant, $\Delta_\text{c}=0$).  The frequency $\Omega$ is lowered during time $\tau_\text{hot}$  by changing the cavity fields from the high initial value $N_\text{c,h}$ to the lower value $N_\text{c,l}$.

\noindent \textit{Step 2:} The temperature of the bath is reduced to $T_\text{eff}$ by detuning the control laser to $\Delta_\text{c,l}$. The frequency $\Omega$ is kept constant. 

\noindent \textit{Step 3:} The particle interacts with a  bath at constant temperature $T_\text{eff}$ via the coupling $\gamma_{\text{eff}}$.  The frequency $\Omega$ is increased to its initial value during time $\tau_\text{cold}$  by enhancing the cavity fields from $N_\text{c,l}$ to $N_{\text{c},\text{h}_\Delta}$. The detuning $\Delta_\text{c}$ is adjusted to keep $\gamma_{\text{opt}}$ constant.

\noindent \textit{Step 4:} In the last isochoric step  all the control parameters are switched back to their initial values.

\begin{figure}[t]
\includegraphics[width=0.49\textwidth]{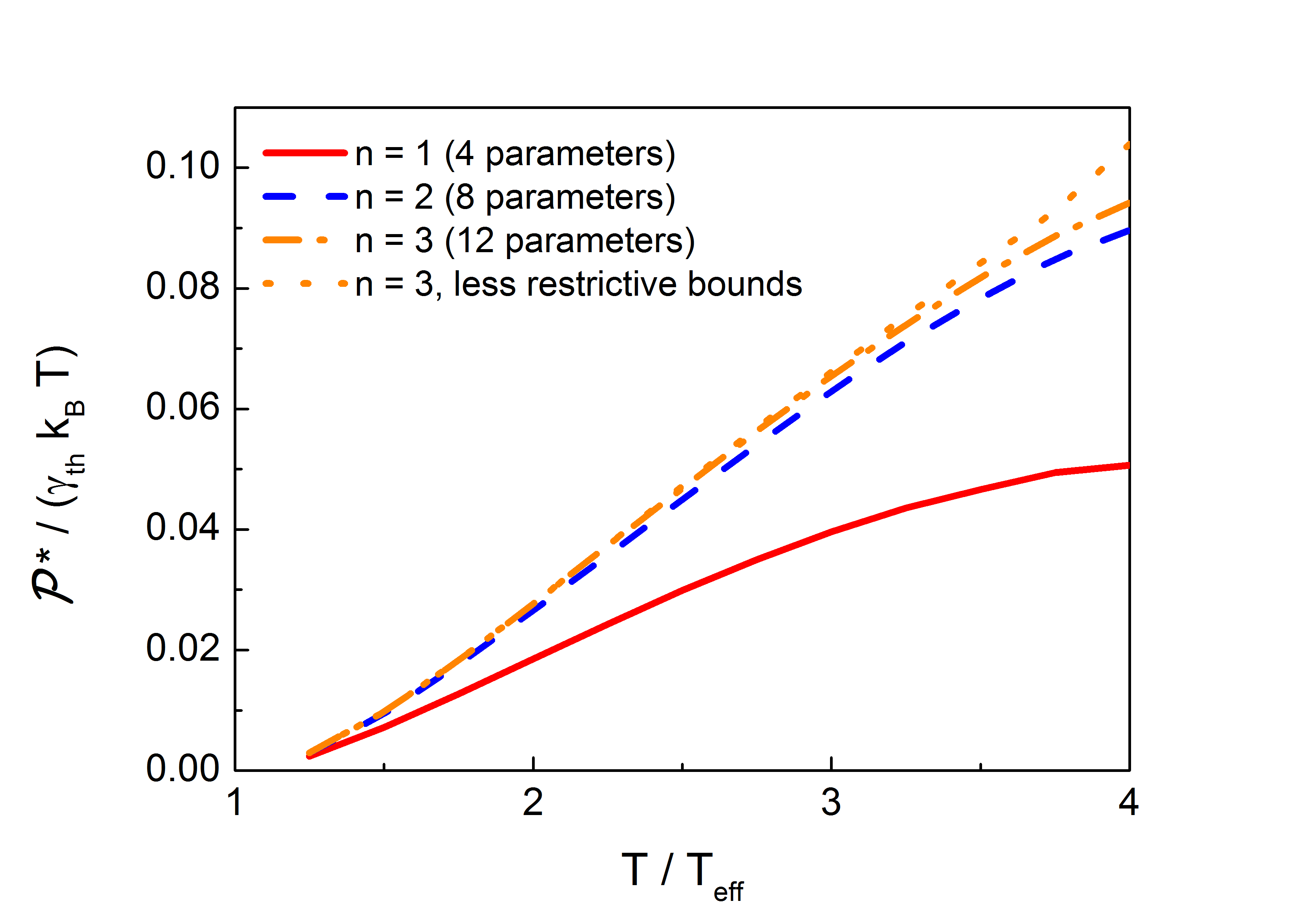}
\includegraphics[width=0.49\textwidth]{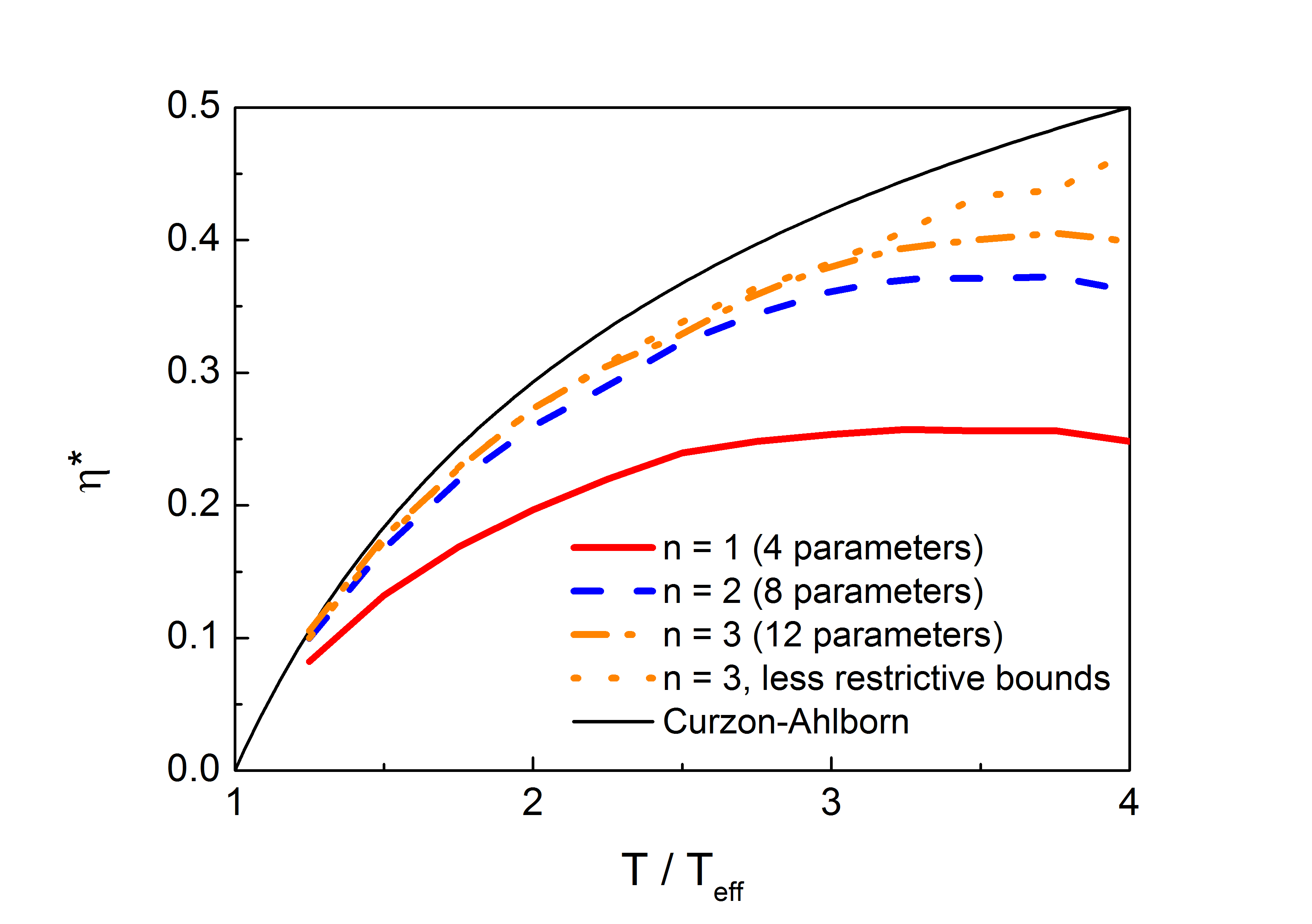}
\caption{Power $\mathcal{P}^{*}$ and efficiency at maximum power $\eta^{*}$ as a function of the ratio of the  bath temperatures, $T/T_\text{eff}$, for various optimization steps, $n=1, 2$ and $3$. The dotted lines show the performance for $n=3$ without the experimental constraints on frequency indicated in Fig.~1e.}
\label{fig3}
\end{figure}

\begin{figure}[ht]
\includegraphics[width=0.49\textwidth]{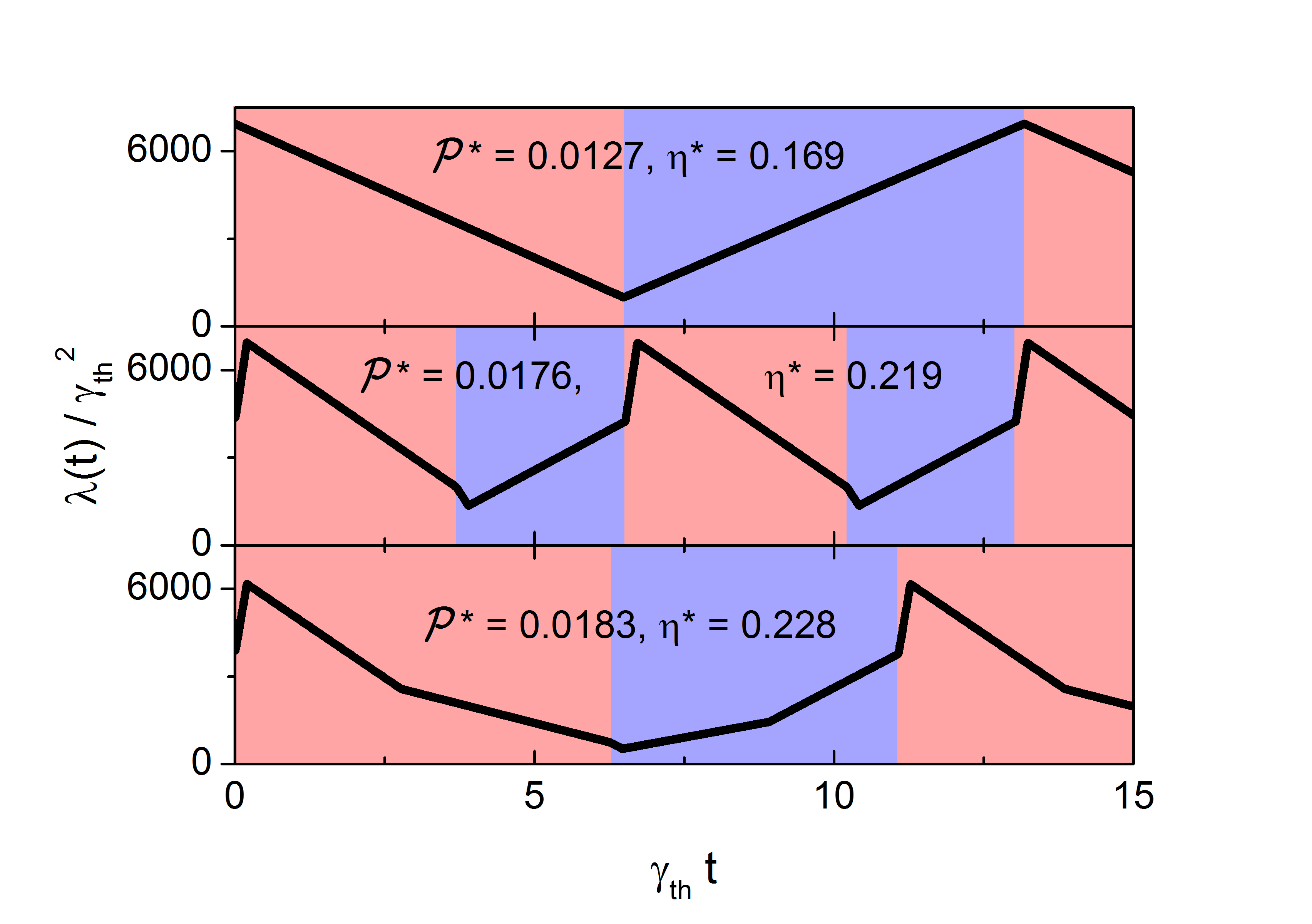}
\caption{Optimal driving protocols $\lambda(t)$ for a fixed temperature ratio $T/T_\text{eff} = 1.75$ and the same optimization steps as in Fig. 2  (top to bottom). The red (blue) shaded regions denote coupling to the hot (cold) bath, the corresponding power output $\mathcal{P}^{*}$ (in units of $\gamma_\text{th} T$) and efficiency $\eta^{*}$ are also stated. Fast frequency variations occur at the transitions between hot and cold baths for higher-order optimization.}
\label{fig2}
\end{figure}

The above cooling-heating sequence based on sideband cooling is illustrated in Fig.~\ref{fig:cycle}c. The thermodynamic Stirling cycle is shown in Fig.~\ref{fig:cycle}e for the theoretical parameters $(\Omega, T_\text{eff})$, and in Fig.~\ref{fig:cycle}d for  the experimental parameters  $(N_\text{c}, \Delta_\text{c})$. A summary of the values used in the  simulations (see Figs.~\ref{fig3} and \ref{fig2}) is given in Tab. 1.

\textit{Optimal protocols.} We shall next determine the driving protocol that maximizes the power output of the  engine.
We begin by writing  the mean heat exchanged  between particle and bath  during a time interval $[t,t+\tau]$ \cite{sek10,Seifert2012},
\begin{align}
Q = \gamma_\text{th} k_B T \tau - m \gamma_\text{eff} \int_{t}^{t+\tau} \text{d}t' \ \sigma_{v}(t'), \label{eq:heattransfer}
\end{align}
where $\sigma_{v}(t) = \langle v^2(t) \rangle$ is the mean-square velocity of the particle and $\gamma_\text{eff} = \gamma_\text{th} + \gamma_\text{opt}$. The work done by the engine during a  full cycle is $-W = Q_{\text{hot}} + Q_{\text{cold}}$. The corresponding power and efficiency 
are accordingly,
\begin{align}
\mathcal{P} =  \frac{Q_{\text{hot}} + Q_{\text{cold}}}{\tau_\text{hot} + \tau_\text{cold}}, \quad \eta = 1 + \frac{Q_{\text{cold}}}{Q_{\text{hot}}} .
\end{align}
In order to compute the above quantities, we  need to evaluate the dynamics of $\sigma_{v}(t)$ in Eq.~\eqref{eq:heattransfer}. Multiplying the Langevin equation  \eqref{eq:oscillator} by $x$, respectively $v$, and taking the ensemble average, we obtain the two equations,
\begin{subequations}
\begin{equation}
\dot{\sigma}_{v} + 2 \gamma_\text{eff} \sigma_{v} + \lambda \dot{\sigma}_{x} = \frac{2 \gamma_\text{eff} k_B T_\text{eff}}{m},  \label{eq:variancedynamicsa} \\
\end{equation}
\begin{equation}
\ddot{\sigma}_{x} + \gamma_\text{eff} \dot{\sigma}_{x} + 2 \lambda \sigma_x - 2 \sigma_v = 0, \label{eq:variancedynamicsb}
\end{equation}
\end{subequations}
where $\sigma_x = \langle x^2(t) \rangle$ denotes the mean-square displacement  and $\lambda = \Omega^2$ the  square frequency of the oscillator. We choose the latter as the control parameter that we wish to  determine  such as to maximize the power  $\mathcal{P}$. The steady state solutions of Eqs.~\eqref{eq:variancedynamicsa} and \eqref{eq:variancedynamicsb} are given by $\sigma_v = \lambda \sigma_x = k_B T_\text{eff}/m$ corresponding to equipartition at temperature $T_\text{eff}$. In the overdamped limit, $\gamma_\text{eff} \gg \Omega$, the velocity thermalizes quasi instantaneously and the dynamics can be described in terms of the slow position variable only. In this regime, the optimal protocol $\lambda(t)$ can be obtained analytically \cite{sch08}. By contrast, in the underdamped limit, $\gamma_\text{eff} \ll \Omega$, the search for the optimal protocol requires solving a set of coupled, nonlinear differential  equations with periodic boundary conditions, a task which is daunting even numerically. 

\begin{table}
\begin{tabular}{|c|c|c|c|c|}
\hline 
cycle step & $T_\text{eff}$ & $\Omega/2\pi$ & $\Delta_\text{c}/2\pi$ & $N_\text{c}/10^8$ \\ 
\hline 
4 $\rightarrow$ 1 & 293 K & 600 kHz & 0 kHz & 21.0 \\ 
\hline 
1 $\rightarrow$ 2 & 293 K & 150 kHz & 0 kHz & 1.31 \\ 
\hline 
2 $\rightarrow$ 3 & 167 K & 150 kHz & 98.7 kHz & 1.32 \\ 
\hline 
3 $\rightarrow$ 4 & 167 K & 600 kHz & 398 kHz & 21.3 \\ 
\hline 
\end{tabular} 
\caption{Values of the control parameters $(\Omega, T_\text{eff})$ of the heat engine and $(P_t, \Delta_c)$ of the optomechanical systems used in the numerical simulations. The parameters for the cycle are the values just before the individual steps. Further experimental parameters are: $T = 300$ K, $\gamma_\text{th}/2\pi = 7.2$ kHz, $\gamma_\text{opt}/2\pi = 5.4$ kHz. The boundaries we set for the mechanical frequencies ($\Omega_\text{max}/2\pi = 600$ kHz, $\Omega_\text{min}/2\pi = 150$ kHz) do not yet restrict the optimal protocol for the temperatures given in the table.} 
\label{tab1}
\end{table}

We tackle this theoretical challenge by introducing piecewise linear trial protocols $\lambda(t)$ with $n=1, 2, 3, ...$ linear segments  that we optimize numerically. We obtain in such a way  a systematic expansion which allows us to specify the optimal protocol  to any desired accuracy. The limit $n\rightarrow \infty$ corresponds to the true optimal protocol. As we will show, the expansion fortunatelly converges rapidly to a stable solution (see Fig.~\ref{fig3}). The first term in the expansion, $n=1$, is a linear change of $\lambda(t)$,  during the coupling to both  hot and cold baths: 
\begin{align}
\lambda(t) = \left\lbrace \begin{array}{ll}
\frac{\lambda_2-\lambda_1}{\tau_\text{hot}} t + \lambda_1, \quad 0 < t < \tau_\text{hot} \\
\frac{\lambda_1-\lambda_2}{\tau_\text{cold}} (t-\tau_\text{hot}) + \lambda_2, \quad \tau_\text{hot} < t < \tau_\text{hot} + \tau_\text{cold}. 
\end{array} \right. \label{eq:protocol-linear}
\end{align}
This simple linear protocol  depends on the four parameters $\tau_\text{hot}$, $\tau_\text{cold}$, $\lambda_1$ and $\lambda_2$.
We find the maximum power $\mathcal{P}^{*}$ and the corresponding efficiency $\eta^{*}$
by numerically solving equations \eqref{eq:variancedynamicsa} and \eqref{eq:variancedynamicsb} for this protocol, computing the corresponding power output and optimizing with respect to the four parameters, keeping the ratio of the temperatures of the two baths, $T/T_\text{eff}$, fixed. The details of the integration and optimization procedure are provided in the Supplementary Material.
For the second term, $n=2$,  we subdivide each of the two linear pieces of the protocol into two parts,
giving  a total of eight parameters to optimize.  We may continue this systematic expansion by each time subdividing a linear segment  into two parts. Figure \ref{fig3} shows the maximum power $\mathcal{P}^{*}$ and the corresponding efficiency $\eta^{*}$
as a function of the temperature ratio $T/T_\text{eff}$, for $n=1, 2$ and $3$. We have performed the numerical optimization by taking the experimentally accessible range of parameters displayed in Fig.~\ref{fig:cycle}e into account. We observe that both power and efficiency at maximum power are significantly improved when going from $n=1$ (four parameters) to $n=2$ (eight parameters). However, the performance of the engine is only slightly enhanced by adding an additional term (12 parameters), indicating that the expansion converges quickly \cite{rem}. Interestingly, the efficiency at maximum power is bounded by the Curzon-Ahlborn efficiency, $\eta_\text{ca}= 1-\sqrt{T_\text{eff}/T}$ \cite{cur75}, which it approaches for small temperature differences. It is also worth to mention the influence of the experimental restrictions we put on the maximum and minimum values $\Omega_\text{max}$, $\Omega_\text{min}$ of the frequency $\Omega_0$. Since larger values of the frequency generally lead to higher power and efficiency, as they allow for a better control of the dynamics, an optimization without these experimental constraints results in improved performance for larger temperature differences (see Fig.~\ref{fig3}).

The optimal protocols found for $n=1, 2$ and $3$ are shown in Fig.~\ref{fig2} for the parameters given in Tab.~\ref{tab1} . We note again a substantial difference between $n=1$ and $n=2$, and minor changes when going to higher orders. The first modification is that the cycle time (indicated by the shaded colored regions) is reduced when more free optimization parameters are available (shorter cycles lead to higher power). The coupling times to hot and cold baths are in general not equal, since   $\gamma_\text{eff} > \gamma_\text{th}$. The second, more fundamental, difference is the appearance of fast variations of the frequency at the transitions between hot and cold baths. Discountinuities in the optimal driving protocol were predicted in the overdamped regime \cite{sch08}. These jumps are interesting, since they are absent in a linear response approach, and their occurrence is therefore a hallmark of far-from-equilibrium behavior.   From a physical point of view, fast frequency variations permit an almost instantaneous change of the velocity of the particle, and hence reduce dissipation \cite{gom08}. In the experimental optomechanical system, these jumps can only be realized in an approximate manner.
If the relative rate of change in the frequency, $\dot{\Omega}/\Omega$, becomes larger than the cavity decay rate, transients in the cavity field have to be taken into account. Nevertheless, our numerical analysis with realistic parameters clearly indicates that signatures of these discontinuities should be experimentally observable and that their essential role in enhancing the performance of the heat engine should be testable.

\textit{Conclusions.} We have introduced a concrete experimental scheme for the realization of an all-optical heat engine in the underdamped regime. We have further developed an efficient optimization procedure that allows to determine the optimal driving protocols to any desired accuracy in a systematic manner. We have performed detailed numerical simulations of the stochastic engine using realistic parameters. We have evaluated the power and the efficiency at maximum power for various optimization steps and specified the corresponding optimal protocols. We have finally discussed the occurrence of frequency jumps whose signature may be observed experimentally. As we have shown, levitated cavity optomechanics is a powerful novel tool for the study of far-from-equilibrium thermodynamics in the underdamped regime. An extension of our analysis to the full quantum mechanical case is possible, accompanying strong experimental efforts to push levitated systems to operate in the quantum regime. 

This work was partially supported by the EU Collaborative Project TherMiQ (Grant Agreement 618074) and the COST Action MP1209.


\begin{thebibliography}{99}
\bibitem{rem} We have optimized with up to 20 parameters ($n=5$) and did not observe any serious improvement compared to $n=3$ (12 parameters).
\end{thebibliography}
\end{document}


\title{An All-Optical Nanomechanical Heat Engine - Supplemental Material}

\author{Andreas Dechant}
\author{Nikolai Kiesel}
\author{Eric Lutz}

\maketitle

In order to find the optimal protocols for the heat engine described in the main text, we first have to compute the dynamics of the variances of velocity $\sigma_v(t)$ and position $\sigma_x(t)$ for a time-dependent spring constant $\lambda(t) = \omega^2(t)$.
This is determined by Eqs.~(4a) and (4b) of the main text.
While an analytical solution is possible for specific types of protocols in the highly underdamped limit $\omega \gg \gamma_\text{eff}$, we instead opt for a numerical solution, as we want to treat the generic case with finite damping.
The numerical solution of the equations of motion for the moments is done by choosing viable initial conditions $\sigma_v(0)$, $\sigma_x(0)$ and $\dot{\sigma}_x(0)$ and then numerically integrating the equations with a given protocol until the solution converges to a set accuracy.
This corresponds to starting the heat engine in a certain state and running it until it operates in a periodic manner.
This process yields the power output and efficiency for a given protocol.
To optimize the protocol, we vary the parameters -- the values of the spring constant and the corresponding coupling times -- within certain boundaries and maximize the power output using the NLopt Fortran library \cite{NLopt}.
The boundaries are set by the limits imposed by the experimental setup in terms of maximal and minimal frequency for the desired temperature difference and the maximal rate of change for the frequency.

More precisely, the integration of the equations of motion is done using a four-step Runge-Kutta method \cite{pre01} until the total integral of the velocity variance $\sigma_v$ changes by less than $10^{-6}$ of its current value after one cycle of the heat engine.
The optimization is done in three steps: First a global optimization is run on the entire available parameter space using a controlled random search algorithm \cite{kae06} using a relative accuracy target of $10^{-4}$ for the power.
As it turns out, the stochastic nature of the optimization algorithm means that it may not always find the true maximal power, especially at this rather large tolerance.
To mitigate this fact, we run several instances of the optimization in parallel and pick best one of those while monitoring the standard deviation of the individual power values to ensure that we are not too far from the true maximum.
Second, we perform another global optimization with a tighter accuracy target of $10^{-5}$ within a limited parameter space around the initial "optimal" parameters.
Third, we use a local optimization algorithm (Constrained Optimization BY Linear Approximations \cite{pow98}) with a relative accuracy target of $10^{-6}$ in order to further improve the result.
The two latter optimization steps also use a smaller accuracy target and time step for the numerical solution of the differential equations.

Using this three-step optimization procedure, the results for the maximal power and corresponding efficiency generally vary by less than $10^{-2}$ of the respective value.
However, the resulting optimal protocols may still differ somewhat between individual optimizations.
This is due to the fact that in the completely underdamped limit $\omega/\gamma_\text{eff} \rightarrow \infty$, the power output and thus the efficiency at maximum power is actually invariant under a change of the total cycle time and thus there exists an approximate additional degree of freedom in the choice of the protocol.